\documentclass[12pt]{article}

\usepackage{a4,graphics,graphicx,amsmath,amssymb,cite}
\usepackage[verbose]{wrapfig}
\catcode`@=11 \@addtoreset{equation}{section} \catcode`@=12

\begin{document}
\title{
{\baselineskip -.2in
%\vbox{\small\hskip 4in \hbox{hep-th/07-10~~~~~~~~~~}}
\vbox{\small\hskip 4in \hbox{IITM/PH/TH/2011/8}}
%\vbox{\small\hskip 4in \hbox{TIFR/TH/07-XX~~~~~~~~}}
} 
\vskip .4in
\vbox{
{\bf \LARGE Non-Supersymmetric Stringy Attractors}
}
%\narrowtext
\author{Pramod Dominic \thanks{email: pramod@physics.iitm.ac.in}~ and
Prasanta K. Tripathy\thanks{email: prasanta@physics.iitm.ac.in}\\
\normalsize{\it  Department of Physics,}\\
\normalsize{\it Indian Institute of Technology Madras,} \\
\normalsize{\it Chennai 600 036, India.}
}}
\maketitle
\begin{abstract}
In this paper we examine the stability of non-supersymmetric attractors in 
type $IIA$ supergravity compactified on a Calabi-Yau manifold, in the presence of sub-leading 
corrections to the $N=2$ pre-potential. We study black hole configurations carrying $D0-D6$ and 
$D0-D4$ charges. We consider the $O(1)$ corrections to the pre-potential given by the Euler
number of the Calabi-Yau manifold. We argue that such corrections in general can not lift the 
zero modes for the $D0-D6$ attractors. However, for the attractors carrying the $D0-D4$ 
charges, they affect the zero modes in the vector multiplet sector. We show that, in the 
presence of such $O(1)$ corrections, the $D0-D4$ attractors can either be stable or unstable 
depending on the geometry of the underlying Calabi-Yau manifold, and on the specific values 
of the charges they carry. 
\end{abstract}
\newpage
\section{Introduction}

The static, spherically symmetric, extremal black holes in any $N=2$ supergravity theory 
in four dimensions, coupled 
to a number of vector multiplets  posses a novel feature \cite{Ferrara:1995ih} which goes with the
name ``attractor mechanism''. In such theories, the scalar fields, which can take arbitrary values at 
spatial infinity, run into a fixed point at the horizon of the black hole, and hence, the black hole 
horizon plays the role of an attractor for these scalar fields. The values of the scalar fields at the 
fixed point is completely determined by the electric and magnetic charges of the black hole. This 
is the reason the attractor mechanism has played a significant role in understanding the 
macroscopic entropy of extremal black holes in supergravity theories. Especially, it explains why 
the macroscopic entropy of the black hole depends only on the quantized gauge charges and not 
on the asymptotic values of the scalar fields.

The attractor behavior depends only on the extremality of the black hole \cite{Ferrara:1997tw,
Goldstein:2005hq}. Hence there can in general exist supersymmetric as well as 
non-supersymmetric attractors in $N=2$ supergravity theories. The supersymmetric attractors 
including the sub-leading corrections have already been investigated in great detail and an 
extensive study has been carried out in understanding the microscopic origin of their entropy
\cite{Strominger:1996kf,Ferrara:1996dd,Ferrara:1996um,Behrndt:1996jn,Behrndt:1998eq,
LopesCardoso:1998wt,Ooguri:2004zv,Strominger:1996sh,David:2002wn,Mohaupt:2008gt}. 
The non-supersymmetric attractors are being explored in recent times 
\cite{Bellucci:2007ds,Bellucci:2008jq}. They are pretty
much similar to their supersymmetric counterparts. Especially, for a class of non-supersymmetric 
black holes, there exists a first order formalism giving rise to a ``fake superpotential'' whose extrema 
give rise to the respective attractors \cite{Ceresole:2007wx}. 
However there are certain important differences as well.  For
example, one of the interesting features the non-supersymmetric attractors possess, in the context 
of $N=2$ theories arising from the compactification of ten dimensional type II supergravity on a 
Calabi-Yau manifold, is the existence of additional number of flat directions in the vector multiplet
sector \cite{Tripathy:2005qp,Nampuri:2007gv}. In contrast, for the supersymmetric attractors at the 
leading order, though the hypermultiplet  moduli are completely decoupled,  the vector multiplet 
moduli are all uniquely fixed at the horizon in terms of the black hole charges. Because of the 
above reason, the supersymmetric attractors are always stable. 
The non-supersymmetric attractors are also stable at the leading 
order. However,  sub-leading corrections might effect the additional flat directions in the vector 
multiplet  sector in the later case. 

One class of corrections which are of particular interest in this context are the sub-leading 
corrections to the $N=2$ prepotential \cite{Cecotti:1988qn,Hosono:1993qy,Hosono:1994ax,Hosono:1995bm} .
%  $$F = D_{abc} \frac{X^aX^bX^c}{X^0} . $$ 
In the case of type $IIA$ compactification on Calabi-Yau three folds, these terms arise from 
the $\alpha'$ corrections to the ten dimensional supergravity Lagrangian \cite{Becker:2002nn}. 
They are of the form:
\begin{equation}
F=D_{abc}\frac{X^aX^bX^c}{X^0}+\alpha_{0a}X^aX^0+i\beta (X^0)^2 \ ,
\label{prepocor}
\end{equation}
where the coefficients $D_{abc},\alpha_{0a}$ and $\beta$ are determined by the topology of the 
underlying Calabi-Yau manifold. Note that, the $\beta$ dependent
term gives rise to $O(1)$ contribution to the above pre-potential, and hence, in the large volume 
limit, it would be natural to first ignore this term and study the behavior of the attractor keeping 
the first two terms in the pre-potential (\ref{prepocor}). For some specific black hole configurations,
this has been carried out in our earlier work \cite{Dominic:2011iz,Dominic:2010yv},
 by solving the attractor conditions explicitly on the 
horizon. We found that the number of zero modes remain unchanged by such corrections. 
The same result has been derived independently  in \cite{Bellucci:2010zd} in a much more 
elegant fashion, using group theoretic techniques.

From the analysis of \cite{Bellucci:2010zd} it is clear that in the absence of the $\beta$-term, the 
symplectic invariance
remains intact and hence by making a symplectic transformation, one can reabsorb the second 
term in Eq.(\ref{prepocor}) in the first one. This is the reason the zero modes remain unchanged.
However, in the presence of the $\beta$-term, the symplectic invariance is broken and the 
geometry of the Calabi-Yau manifold is corrected by this term. Thus there is a possibility of lifting the 
zero modes in presence of such corrections. Such corrections, for one and two parameter models 
has already been considered in \cite{Bellucci:2007eh,Bellucci:2008tx}, and, as expected, the flat 
directions are lifted by them. In the present work we will generalize these results to arbitrary 
$n$-parameter models.

The plan of the paper is as follows. In the next section we will discuss some of the preliminaries 
of the attractor mechanism for non-supersymmetric black holes. In \S3 we will discuss the $D0-D6$
configuration and argue that the $\beta$-term does not effect the number of zero modes. \S4
analyzes the $D0-D4$ configuration. In this case, we will show that, for suitable charge 
configurations, the sub-leading corrections in Eq.(\ref{prepocor}) can in fact lift all the zero modes
leading to a stable attractor. Finally in \S5 we will summarize the results and discuss some of 
the future prospects. Some of the computational details will be carried out in the appendices. 

\section{Preliminaries} 

In this section we will review the non-supersymmetric attractors and recapitulate  some of the recent 
developments on their stability that will be relevant for the subsequent discussions. Many of the results 
we summarize in this section were originally derived in \cite{Ferrara:1997tw}. We will closely follow their 
notations in most of our discussions in this section.

Throughout this paper, we 
will restrict ourselves to the $N=2$ theories arising from the compactification of type IIA 
supergravity on a Calabi-Yau manifold. Furthermore, we will ignore the hypermultiplet 
sector of the moduli space entirely since they do not play any role here. The bosonic part 
of the corresponding supergravity action coupled to $n$ vector multiplets is given by
:
\begin{eqnarray}
S = \int d^4x \sqrt{-g}\left( - \frac{1}{2} R 
+  g_{i\bar j} \partial_\mu x^i\partial^\mu\bar x^{\bar j} %h^{\mu\nu}  
-  \mu_{ab} {\cal F}^a_{\mu\nu}{\cal F}^{b \mu\nu}
% h^{\mu\lambda}h^{\nu\rho}  \right.\nonumber\\ &- &\left.
-  \nu_{ab} {\cal F}^a_{\mu\nu} *{\cal F}^{b \mu\nu}
%h^{\mu\lambda}h^{\nu\rho}
\right) \ .
\label{sugraaction}
\end{eqnarray}
Here $x^i \ (i=1,\cdots,n)$ are the complex scalars parametrizing the vector multiplet moduli space, 
$g_{i\bar j}$ is the metric on the moduli space and $x^\mu \ (\mu = 0,\cdots, 3)$ are the space-time 
co-ordinates. The space-time metric is denoted by $g_{\mu\nu}$ with determinant $g$. ${\cal F}^a$ 
are the field strengths corresponding to the $(n+1)$ gauge fields ${\cal A}^a$. The gauge couplings 
$\mu_{ab}$ and $\nu_{ab}$ are completely determined by the $N=2$ pre-potential $F$. 

In this paper we will study static, spherically symmetric black holes. For such black holes,
 the space-time metric %$h_{\mu\nu}$ 
 takes the form:
\begin{eqnarray}
ds^2 %= h_{\mu\nu} dx^\mu dx^\nu 
= e^{2 U} dt^2 - e^{-2U} \gamma_{mn} dx^mdx^n \ .
\label{metricansz}
\end{eqnarray}
We further consider intersecting brane configurations with $D0$ branes carrying a net charge $q_0
$ and $p^0$ number of $D6$ branes wrapping the Calabi-Yau manifold ${\cal M}$. In addition, 
there are $q_i$ number of $D2$ branes wrapping the two cycles $\Sigma_i$ and $p^i$ number 
of $D4$ branes wrapping the four cycles dual to $\Sigma_i$.  The gauge field strength for such
a configuration is
\begin{eqnarray}
{\cal F}^{a} = e^{2 U}\ \frac{q^a}{r^2}\ dt\wedge dr - p^a \sin\theta\ d \theta \wedge d \phi \ .
\label{gfieldstr}
\end{eqnarray}

Substituting the above expression for the gauge field strength and the metric ansatz Eq.(\ref{metricansz}) 
in the supergravity action (\ref{sugraaction}) and integrating out the angular variables we get 
an effective one dimensional theory with a potential
\begin{eqnarray}
V = e^K\left(g^{i\bar j} \nabla_iW (\nabla_jW)^* + |W|^2\right) \ . \label{effectivepot}  
\end{eqnarray}
Here $g_{i\bar j} = \partial_i\partial_{\bar j} K$ is the moduli space metric with its inverse 
$g^{i\bar j}$, and $\nabla_iW = \partial_iW + \partial_i K W$. The K\"ahler potential $K$ 
and superpotential $W$ are determined in terms of the $N=2$ prepotential $F$ as 
\begin{eqnarray} \label{kahler}
K &=& - \ln\left(\Im\left(\sum_{a=0}^n\bar X^a\partial_aF\right)\right)  \ , \\ \label{superpot}
W &=& \sum_{a=0}^n\left( q_a X^a - p^a\partial_aF\right) \ .
\end{eqnarray}

The attractor point is given by the critical values of the effective potential $\partial_i V = 0$.
We have stable attractors if, in addition,  the matrix of second derivatives of $V$ is positive 
definite \cite{Goldstein:2005hq}.

For type $IIA$ compactification on a Calabi-Yau manifold, in the large volume limit the leading order 
term of the pre-potential is given by 
\begin{eqnarray}
F = D_{abc} \frac{X^a X^b X^c}{X^0} \ .
\label{leadingprep}
\end{eqnarray}
The non-supersymmetric attractors with the above pre-potential has been studied in 
Ref.\cite{Tripathy:2005qp}. In this paper we are interested in studying $D0-D4$ and $D0-D6$ 
configurations. Let us first consider the $D0-D4$ configuration with the above pre-potential. 
The critical points of the effective black hole potential (\ref{effectivepot}) are given by
\begin{eqnarray}
x^a_s =  i p^a \sqrt{ \frac{q_0}{D}} \  \  {\rm and} \ \  x^a_{ns} = i p^a \sqrt{- \frac{q_0}{D}} \ .
\end{eqnarray}
Here $D = D_{abc} p^a p^b p^c$. The critical point $x^a_s$ exists when $q_0D > 0$ and corresponds 
to the supersymmetric attractor. The black hole entropy in this case is given by $S = 2\pi \sqrt{q_0 D}$.
For $q_0D<0$ the critical point is given by $x^a_{ns}$ and this corresponds to the non-supersymmetric
black hole with entropy $S = 2\pi \sqrt{-q_0 D}$. The supersymmetric solution is perturbatively stable. 
However a priori there is no  reason 
why the non-supersymmetric solution should be stable and hence we need to explicitly verify its 
stability  by computing the corresponding mass matrix. 

The mass matrix for $D0-D4$ system has been computed in \cite{Tripathy:2005qp}. It has the form
\begin{eqnarray}
M = 32\  q_0^2 \ e^{K_0}\ \left(
\begin{matrix} g_{a\bar b}^0 & 0 \cr 0 & - (9D/4q_0) D_a D_b
\end{matrix}
\right)\ .
\end{eqnarray}
Here $K_0$ and $ g^0_{a\bar b}$ are the K\"ahler form and the moduli space metric evaluated at the 
attractor point. Clearly, the matrix $M$ has $(n+1)$ positive and and $(n-1)$ zero  eigenvalues and 
hence, to the leading order, the effective potential has  $(n+1)$ stable directions and $(n-1)$ flat 
directions.

For the $D0-D6$ configuration supersymmetric solution does not exist. The non-supersymmetric 
critical point is given by \cite{Nampuri:2007gv}:
\begin{eqnarray}
x^a_{ns} = \left\{
\begin{matrix} 
i \hat x^a_- \left(\frac{q_0}{p^0}\right)^{(1/3)} \ {\rm For } \ q_0p^0>0 , \cr
i \hat x^a_+ \left(-\frac{q_0}{p^0}\right)^{(1/3)} \ {\rm For } \ q_0p^0 < 0 \ ,
\end{matrix}\right.
\end{eqnarray}
where $\hat x^a_+$ and $\hat x^a_-$ are two arbitrary real vectors confined to the 
hypersurfaces $D_{abc}\hat x^a_+
\hat x^b_+\hat x^c_+ = +1$ and $D_{abc} \hat x^a_-\hat x^b_-\hat x^c_-=-1$ respectively. 
Since this defines $(n-1)$ dimensional hypersurface in a $2n$ dimensional space, we have
$(n-1)$ flat directions in this case as well. 

It would be natural to ask if these flat directions, for both $D0-D4$ as well as $D0-D6$ systems
can be lifted by considering sub-leading corrections to the pre-potential (\ref{leadingprep}).
This has been partially carried out in Ref. \cite{Dominic:2011iz} by considering the next sub-leading
term in the pre-potential:
\begin{eqnarray}
F=D_{abc}\frac{X^aX^bX^c}{X^0}+\alpha_{0a}X^aX^0 \ .
\label{classicalgeom}
\end{eqnarray}
For the $D0-D4$ system the only effect of this correction is to shift the $D0$ charge. For the 
$D0-D6$ system the computation becomes much more involved. However the qualitative 
behavior of the solution remains unchanged. The correction term only deforms the moduli 
space without changing its dimensionality. 

In this paper we will consider the most general pre-potential which incorporates the perturbative
corrections to all orders. We will see that the $D0-D6$ system still admits flat directions where as
the $D0-D4$ system could be completely stabilized. For a two parameter model, this has been 
carried out in Ref.\cite{Bellucci:2008tx}. Our goal here is to generalize this result to a general 
$n$-parameter model.

\section{The $D0-D6$ system}

In this section we will study the non-supersymmetric attractor for the $D0-D6$ system with perturbative 
corrections. The pre-potential which incorporates perturbative corrections to all orders is given by
\cite{Hosono:1993qy,Hosono:1994ax,Hosono:1995bm}:
\begin{eqnarray}
F=D_{abc}\frac{X^aX^bX^c}{X^0}+\alpha_{0a}X^aX^0+i\beta (X^0)^2 \ .
\label{fullprep}
\end{eqnarray}
Here $D_{abc} = (1/3!) \int_{\cal M} J_a\wedge J_b\wedge J_c $ are the triple intersection 
numbers with $J_a$s being two-forms on the Calabi-Yau manifold ${\cal M}$ belonging to the 
cohomology $H^2({{\cal M},\mathbb Z})$, where as  $\alpha_{0a} = - (1/24) \int_{\cal M} c_2\wedge J_a$ 
and $\beta = - \zeta(3)\chi/(16\pi^3)$ with $c_2$ and $\chi$ being the second Chern class and  the Euler 
number of ${\cal M}$ respectively.

Unlike the pre-potential (\ref{classicalgeom}), the term depending on $\beta$ in Eq.(\ref{fullprep}) actually
modifies the classical geometry of the Calabi-Yau manifold ${\cal M}$.  The corresponding K\"ahler 
potential can be found to have the form
\begin{equation} \label{kahlerpotful}
K=-\ln\left(- i M - 4\beta \right) \ , 
\end{equation}
where $M=D_{abc}(x^{a}-{\bar{x}}^{a})(x^{b}-{\bar{x}}^{b})(x^{c}-{\bar{x}}^{c})$. % and $L=4\beta$.
Here we have used the formula (\ref{kahler}) for the K\"ahler potential, introduced the notation 
$x^a = X^a/X^0$ and subsequently chosen the gauge $X^0=1$. The corresponding metric 
$g_{a\bar b} = \partial_a\partial_{\bar b} K$ can easily be calculated. We have 
\begin{eqnarray} \label{metric}
g_{a\bar b} = {3\over {M- 4 i \beta}} \left( 2 M_{ab} - {3\over {M- 4 i \beta}} M_a M_b\right)~,
\end{eqnarray}
and its inverse
\begin{eqnarray}
g^{a\bar b} = {M- 4 i \beta \over 6} \left( M^{ab} - {3\over {M+ 8 i \beta}} (x^a - \bar x^a)
(x^b - \bar x^b) \right) ~.
\end{eqnarray}
For convenience we have introduced $M_{ab} = D_{abc} (x^c - \bar x^c)$ and $M_a = M_{ac}(x^c - \bar x^c)$.
$M^{ab}$ is the inverse of the matrix $M_{ab}$.

%It is useful to have the derivatives of the metric for the future use(where we have used $x^{a}=p^{a}t$).
%\begin{eqnarray}
%\partial_{a}g^{b\bar{c}}&=&-\frac{D(1+s)it_{2}}{3}\left(D^{pb}D^{qc}D_{apq}+\frac{3}{D(1-2s)}\left({\delta_{a}}^{b}p^{c}+{\delta_{a}}^{c}p^{b}\right)\right)\nonumber\\&+&it_{2}D_{a}D^{bc}+\frac{9ist_{2}D_{a}p^{b}p^{c}}{D(1-2s)^{2}}
%\end{eqnarray}

We will now consider the $D0-D6$ system. The superpotential for this system can be derived using 
Eqs.(\ref{fullprep}) and (\ref{superpot}). We find
\begin{equation}
W=q_{0}-2ip^{0}\beta-p^{0}{\alpha_{0}}_{a}x^{a}+p^{0}D_{abc}x^{a}x^{b}x^{c}  \ .
\label{super1}
\end{equation}
To find the attractor point, we need to consider the critical points of the black hole effective potential 
(\ref{effectivepot}). In other words, we need to consider the equations of motion:
\begin{equation}
g^{b\bar c} \nabla_a\nabla_b W \overline{\nabla_cW} + 2 \nabla_a W \overline{W}
+ \partial_a g^{b\bar c} \nabla_b W\overline{\nabla_c W} = 0~.
\end{equation}

Taking a clue from the $D0-D6$ solutions found in Refs.\cite{Nampuri:2007gv,Dominic:2011iz} we set 
the ansatz $x^a = \hat x^a t = \hat x^a (t_1 + i t_2)$ with some real vector $\hat x^a$. Substituting this 
ansatz, the expression for the K\"ahler potential as given in Eq.(\ref{kahlerpotful}) and the superpotential 
(\ref{super1}) in the above we can rewrite the equations of motion in terms of $t_1$ and $t_2$. This 
has been carried out in the appendix. The exact expressions for the equations of motion are lengthy 
and hence we will not reproduce them here. For $\beta=0$ the exact solution has been found  in
Ref.\cite{Dominic:2011iz}. For $\beta \neq 0$ it is not possible to find exact analytical expression for 
$t_1$ and $t_2$. However, in the large charge limit, we can do a perturbative analysis. For specific 
models it is also possible to solve these equations numerically. Thus we will assume that, for 
$\beta\neq 0$, the solution indeed exists and we will denote this solution to be 
$x^a_0 = \hat x^a t_0 =  \hat x^a (t_{01} + i t_{02})$.  From the analysis in  appendix A, it is then clear 
that the $n$ real vectors $\hat x^a$ obey one real constraint of the form
$$ f(\hat x^a, p^0, q_0, D_{abc},\alpha_{0a},\beta) = 0 \ . $$
Thus there is a $(n-1)$ dimensional hypersurface of attractor points in the case of $D0-D6$ system,
and the perturbative corrections do not lift any of the $(n-1)$ flat directions originally existed in the 
leading order result with pre-potential (\ref{leadingprep}). The only effect of the correction terms is 
to deform the hypersurface on which the $\hat x^a$ live and not change its dimensionality.

\section{The D0-D4 system}

In this section we will turn our attention to the more interesting case of a $D0-D4$ system.  Using the 
expression (\ref{fullprep}) for the pre-potential, the superpotential (\ref{superpot}) can be found to take 
the form:
\begin{equation}
W=q_0 - \alpha_{0a} p^a -3D_{ab}x^{a}x^{b} \ . 
\end{equation}
Here we have introduced $D_{ab} = D_{abc} p^c$. Note that the above expression is identical to the 
superpotential corresponding to the leading pre-potential (\ref{leadingprep}),  with a shifted $D0$ 
charge $q = q_0 - \alpha_{0a} p^a$. However, because of the presence of the $\beta$-term in 
Eq.(\ref{fullprep}), the geometry of the Calabi-Yau manifold changes, as can be seen from
the expression for the metric in Eq.(\ref{metric}), and hence the black hole effective potential 
(\ref{effectivepot}) also gets modified non-trivially.

We are interested in studying the stability of non-supersymmetric attractors for the above system. To get
the attractor point, we need to consider solutions to the equation
   \begin{equation}\label{veffprim}
 g^{b\bar c} \nabla_a\nabla_b W \overline{\nabla_cW} + 2 \nabla_a W \overline{W}
+ \partial_a g^{b\bar c} \nabla_b W\overline{\nabla_c W} = 0~,
\end{equation}
such that $\nabla_a W \neq 0$. It is straightforward to evaluate the covariant derivatives in each of the 
terms of the above equation. We will set the ansatz $x^a = p^a t = p^a (t_1 + i t_2)$ in each of these
terms and equate the real and imaginary parts of Eq.(\ref{veffprim}) to zero. After considerable 
simplification we find:
\begin{eqnarray}
\label{real}
0&=& t_1 t_2 \left(D t_2^3-\beta \right) \left(\beta +2 D t_2^3\right) \left(4 D^2 \left(2 t_2^6+3 t_1^2 t_2^4\right)-D \left(4 q t_2^4+\beta 
   \left(3 t_1^2+t_2^2\right) t_2\right)\right.\nonumber\\&+&\left.\beta  \left(2 \beta +q t_2\right)\right) \ ,
   \end{eqnarray}
   \begin{eqnarray}
   \label{imagin}
 0&=&  t_2^4 \beta ^2 \left(3 D^2 \left(15 t_1^4+46 t_2^2 t_1^2+23 t_2^4\right)+6 D q \left(t_2^2-5 t_1^2\right)+5 q^2\right)+4 \left(t_2^2-t_1^2\right) \beta ^4 \nonumber\\
   &+&8 D^2 t_2^{10} \left(D^2 \left(9 t_1^4+4 t_2^2 t_1^2-t_2^4\right)-6 D q
   t_1^2+q^2\right)-4 t_2^3 \beta ^3 \left(D \left(5 t_1^2+t_2^2\right)+3 q\right)\nonumber\\ 
    &+&4 D t_2^7 \beta  \left(D^2 \left(-9 t_1^4+4
   t_2^2 t_1^2+5 t_2^4\right)+6 D q \left(t_1^2-2 t_2^2\right)-q^2\right) \ ,
       \end{eqnarray}
where $D = D_{abc} p^a p^b p^c$. We will look for axion free solution to the above equations. In  this case 
$t_1=0$ and hence the first of the above two equations  is satisfied trivially where as the second equation 
takes the simple form:
\begin{eqnarray}
\label{fact}
0&=&\left(-D t_2^3+q t_2-2 \beta \right) \left(8 D^3 t_2^9+4 D^2 t_2^6 \left(2 q t_2-9 \beta \right)+D t_2^3 \beta  \left(3 \beta -4 q t_2\right)\right.\nonumber\\&+&\left.\beta ^2 \left(5
   q t_2-2 \beta \right)\right)  \ . 
   \end{eqnarray}
   The first factor $\left(-D t_2^3+q t_2-2 \beta \right) = 0$ in the right hand side corresponds to the 
   supersymmetry condition $\nabla_a W =0$. For the non-supersymmetric attractor, we have 
   \begin{eqnarray}
\label{fact1}
0=\left(8 D^3 t_2^9+4 D^2 t_2^6 \left(2 q t_2-9 \beta \right)+D t_2^3 \beta  \left(3 \beta -4 q t_2\right)+\beta ^2 \left(5
   q t_2-2 \beta \right)\right) \ .
     \end{eqnarray}   
 To simplify it further we introduce the following redefinition:
\begin{eqnarray}
t_{2}&=&y\sqrt{-q/D}\\
\beta&=&\gamma q\sqrt{-q/D}
\end{eqnarray}  
In terms of the rescaled variable $y$ and the parameter $\gamma$, Eq.(\ref{fact1}) becomes:
\begin{eqnarray}\label{theequation}
8 y^9-8 y^7+36 y^6 \gamma -4 y^4 \gamma +3 y^3 \gamma ^2-5 y \gamma ^2+2
   \gamma ^3=0
   \end{eqnarray}  
This coincides with the equation for non-supersymmetric attractors in the one parameter 
model. This  equation has been studied extensively in \cite{Bellucci:2007eh}. We will be interested 
in the large charge limit. In this limit, the solution exists and it will depend on the value of $\gamma$.
We will not be interested in the exact value of the solution. Let $y=y_0$ be the solution to 
Eq.(\ref{theequation}) in the large charge limit.  Thus the non-supersymmetric attractor for a 
$n$-parameter model will correspond to  $x^a = i p^a \ y_0\ \sqrt{-q/D} $. In the large charge limit,
$\gamma \ll 1$. Setting $\gamma =0$ we recover the leading solution $y_0=1$. For nonzero $\gamma$
the value of $y_0$ will be different from $1$, however it will still be $O(1)$ for $\gamma \ll 1$.
The correction to the entropy of the black hole is given by
\begin{eqnarray}
S&=& \pi\sqrt{-q D}\   \ \frac{
\left(12 y_0^7-3 y_0^4 \gamma +4 y_0^3-18 y_0^2 \gamma +12 y_0 \gamma ^2+\gamma \right)}
{4 \left(2 y_0^3-\gamma \right) \left(y_0^3+\gamma \right)} \ . 
\end{eqnarray}

We will now analyze the stability of this non-supersymmetric solution. We need to consider the mass matrix
and compute its eigenvalues. For this, it is convenient to introduce the parameter $\omega = \gamma^{1/3}$ 
and the variable $z = y/\omega$. We will evaluate the mass matrix and express it in terms of the solution 
$z_0 = y_0/\omega$. The details of the computation is carried out in appendix B. The mass matrix has 
the expression
\begin{eqnarray*}
M=\left(\frac{3D_aD_d}{D}E_1(z_0)-D_{ad}E_2(z_0)\right)\otimes I
+\left(A_1(z_0)D_{ad}+A_2(z_0)\frac{D_aD_d}{D}\right)\otimes\sigma^3  .  \ \ \ 
\end{eqnarray*}
The functions $E_1(z),E_2(z),A_1(z)$ and $A_2(z)$ are defined in appendix B. The mass matrix $M$ can 
be rewritten in the block diagonal form as 
\begin{eqnarray}\label{blockdiag}
e^{-K_0} M&=& - \frac{24q}{D\left(8z_0^9+36z_0^6+3z_0^3+2\right)}\left(
\begin{array}{cc}
 M_t(z_0)& 0 \\
 0 &M_b(z_0)
\end{array}
\right) \ ,
\end{eqnarray}
where
\begin{eqnarray}
M_t(z)&=&\frac{3D_aD_d\left(16\left(4z^9-2z^6+6z^3+ 7\right)z^6+z^3\right)}{8z^6+4z^3+5}-DD_{ad} \left(4 z^3+1\right)\left(1-2z^3\right)^2 \cr
M_b(z)&= &3\left(DD_{ad}\left(1-2z^3\right)^2+D_aD_d\left(8z^6+4z^3+5\right)z^3\right) \ .
\label{matrixmb}
\end{eqnarray}
Here $K_0$ is the K\"ahler potential evaluated at the attractor point $z=z_0$ and $g^0_{a\bar b}$ is 
the corresponding moduli space metric (at $z=z_0$). We further used the notation $D_a = D_{ab} p^b$.
Note that, since $y_0\sim O(1)$ and $\gamma \ll 1$,  we have $z_0 = y_0/\omega \gg 1$. We can see
that  the matrix $M_t = (64/3) (-Dq) z_0^9 g^0_{a\bar b} + O(z_0^6) $ where as $M_b = 24 z_0^9 D_a D_b + O(z^6_0)$. 
Since we assume that the metric is non-degenerate everywhere in the moduli space, especially at 
the attractor point, the $O(z_0^6)$ term in $M_t$ can't destabilize any of the eigenvalues. Thus for 
our solution the matrix $M_t$ is positive definite. In contrast, the leading term in $M_b$ has the matrix
$D_a D_b$ which has one positive and $(n-1)$ zero eigenvalues. In this case the $O(z_0^6)$
sub-leading terms will play a crucial role in determining the nature of the eigenvalues of $M_b$. 
Keeping terms up to $O(z_0^6)$ in $M_b$ we find 
\begin{eqnarray}\label{leadingmass} 
\frac{M_b}{\left(8z_0^9+36z_0^6+3z_0^3+2\right)} 
=  \frac{3}{2z_0^3} \left( 2  ( z_0^3 -4)D_a D_d + D D_{ad} \right) + O\left(\frac{1}{z_0^6}\right) \ . 
\end{eqnarray} 
The matrix $\left( 2 ( z_0^3 -4)D_a D_d + D D_{ad} \right)$ will in general have non-zero determinant 
except for some very special choices of $D4$-brane charges. Thus, the corrections in Eq.(\ref{fullprep})
will in general lift all the flat directions for the $D0-D4$ system. In the following subsections we will see 
the case of two and three
parameter models, where all the flat directions are lifted for generic $D0-D4$ system. More generally, 
the non-supersymmetric attractor will be stable if the matrix $M_b$ in Eq.(\ref{matrixmb}) is positive 
definite. 

\subsection{Two parameter models}

In order to illustrate that the mass matrix for the $D0-D4$ system will have all non-zero eigenvalues we 
will first consider the simplest scenario of a general two parameter model. In this case  $M_b$ is merely 
a $2\times 2$ matrix and we can diagonalize it quite trivially. It has the eigenvalues 
\begin{eqnarray}\label{eignvlb}
\lambda_1&=&\frac{1}{D}\left(a(z_0)+\sqrt{b(z_0)}\right) \ , \nonumber\\
\lambda_2&=&\frac{1}{D}\left(a(z_0)-\sqrt{b(z_0)}\right) \ ,
\end{eqnarray}
both being non-zero. Here we have defined 
\begin{eqnarray}
a(z)&=&-q\left(8z^9+36z^6+3z^3+2\right)\left(D\left(1-2z^3\right)^2(D_{11}+D_{22})\right.\nonumber\\&+&\left.\left(8 z^6+4 z^3+5\right) z^3
   \left({D_1}^2+{D_2}^2\right)\right)\nonumber\\
b(z)&=&q^2\left(8z^9+36z^6+3z^3+2\right)^2\left(D^2\left(1-2z^3\right)^4 \left((D_{11}-D_{22})^2+4D_{12}^2\right)
\right.\nonumber\\&+&\left.
2D\left(1-2 z^3\right)^2\left(8z^6+4z^3+5\right)z^3\big\{\left({D_1}^2- {D_2}^2\right) (D_{11}-D_{22})\right.
\cr & + & 4D_1D_{12}D_2\big\} +
%\right.\nonumber\\&+&
\left.\left(8z^6+4z^3+5\right)^2 z^6\left((D_1)^2+(D_2)^2\right)^2\right)
\end{eqnarray}
Note that, though we have an overall negative sign in the expression for $a$ in the above equation,
 for the non-supersymmetric case the leading term in $a/D$ contains a positive coefficient times
$-q/D$ and hence $a/D>0$. Assuming $D>0$, we find $\lambda_1 >0$. However, $\lambda_2$ 
need not be positive. It is positive only if $a>\sqrt{b}$, or in other words, if $a^2-b>0$. (For $D<0$,
the role of $\lambda_1$ and $\lambda_2$ are exchanged however the same condition holds.) 
Thus we see that one of the eigenvalues of the matrix $M_b$ is always positive. However, the 
other non-zero eigenvalue is positive only if $a^2 - b>0$. It is straightforward to see that,
\begin{eqnarray}
a^2 - b =4q^2\left(8 z^9+8 z^6+z^3+1\right) \left(16 z^{12}+64 z^9-30 z^6+z^3-2\right)^2
 {\rm det}(D_{ab}) \  , \nonumber
\end{eqnarray}
where ${\rm det}(D_{ab}) = \left(D_{11}D_{22} - D_{12}^2\right)$ is the determinant of the matrix $D_{ab}$.
Clearly, the coefficient of ${\rm det}(D_{ab})$ in the right hand side is positive. 
Thus we have a stable non-supersymmetric attractor only if ${\rm det}(D_{ab})>0$. However, for ${\rm det}
(D_{ab})<0$ one of the directions becomes unstable. We have seen
that the eigenvalues in Eq.(\ref{eignvlb}) are quite complicated. However the condition that they are both
positive is remarkably simply and is  determined by the determinant of the matrix $D_{ab}$. Since
$D_{ab} = D_{abc} p^c$ this condition depends not only on the topology of the Calabi-Yau manifold, but
also on the charges of the $D0-D4$ configuration. For a given Calabi-Yau manifold, as  we move smoothly 
on the charge lattice, the non-supersymmetric attractor transforms from a stable one to an unstable configuration.  
For the first case, i.e. when ${\rm det}(D_{ab})>0$,
our results are in agreement with 
the special case discussed in \cite{Bellucci:2008tx}. Here by considering a general configuration we find 
that the attractor is stable only in a sub-lattice of the charge lattice for the $D0-D4$ system.

\subsection{Three parameter models}

In the previous subsection we have seen that for an arbitrary two parameter Calabi-Yau manifold, the 
non-supersymmetric attractor becomes stable only if the determinant of the matrix $D_{ab}= D_{abc}p^c$
is positive. In this subsection we will analyze the stability conditions in three parameter models. 

The matrix $M_b$ is now a $3\times 3$ matrix and in principle we can diagonalize it by brute force. However
the expressions for the eigenvalues are quite complicated and we can't get any insight by reproducing them
here. In what follows, we will do a first order perturbation to evaluate the eigenvalues. For our purpose it is 
sufficient to consider the leading terms as given in Eq.(\ref{leadingmass}). For convenience we will make a 
change of variables $x_0=1/z_0$. The relevant term in the lower block of the mass matrix (\ref{blockdiag}) is:
\begin{eqnarray} \label{leadm}
 M_{\rm lead} = - 72 \frac{q}{D} D_a D_d - 36 \frac{q}{D} x_0^3 \left(DD_{ab} -8D_{a}D_{b}\right)\ .
\end{eqnarray}
We will now use the first order perturbation theory to evaluate the eigenvalues of the matrix (\ref{leadm}). We 
know that in the three parameter case the tree level part $D_aD_d$ of the matrix $M_{\rm lead}$ has two 
zero modes. We will choose a basis where this matrix is diagonal and the corresponding eigenvectors  
are given by
\begin{eqnarray}
\psi_1&=&\frac{1}{\sqrt{D_1^2 + D_3^2}}
\left\{- D_3,0,D_1\right\} \ , \nonumber\\
\psi_2&=& \frac{1}{\sqrt{\left(D_1^2+D_3^2\right) \left(D_1^2+D_2^2+D_3^2\right)}}
   \left\{-D_1 D_2, D_1^2+D_3^2,-D_2
   D_3 \right\} \ , \nonumber\\
\psi_3&=&\frac{1}{\sqrt{D_1^2+D_2^2+D_3^2}}
\left\{D_1,D_2,D_3
   \right\} \ .
\end{eqnarray}
The first two eigenvectors $\psi_1$ and $\psi_2$ correspond to the zero modes. 
The corrected eigenvalues of the matrix $M_{\rm lead}$ correspond to the eigenvalues of the
following $2\times 2$ matrix:
\begin{eqnarray}
M_p=- \frac{18 \ q \ x^3}{D (D_1^2+D_3^2)}\left(
\begin{array}{cc}
 J_{11} & J_{12} \\
 J_{12} & J_{22}
\end{array}
\right) = - \frac{18 \ q \ x^3}{D (D_1^2+D_3^2)} J  \ , \end{eqnarray}
where the matrix $J$ is defined in terms of its elements $J_{ab}$ such that 
\begin{eqnarray}
J_{11}&=&D \left(D_1^2D_{33}-2D_1D_{13}D_3+D_{11}D_3^2\right) \ , \cr
J_{12}&=&\frac{D}{\sqrt{D_1^2+D_2^2+D_3^2}} 
\Big(D_1^2 \left( D_1 D_{23} - D_3 D_{12} - D_2 D_{13}\right) \cr
&+&  D_3^2 \left(D_1 D_{23} + D_2 D_{13}-D_3
   D_{12}\right) + D_1 D_2 D_3 \left(D_{11}-D_{33}\right) \Big) \ , \cr
J_{22}&=&\frac{D}{D_1^2+D_2^2+D_3^2} \Big( D_{22} \left(D_1^2 + D_3^2\right)^2 
+ D_2^2 \left(D_1^2 D_{11} + D_3^2 D_{33}\right)
\cr &- & 2 D_2 \left(D_1^2 + D_3^2\right) \left(D_1 D_{12} + D_3 D_{23}\right)\Big)\ .
\end{eqnarray}
The eigenvalues of the matrix J are given by
\begin{eqnarray}
\lambda_1&=&\frac{1}{2 \left(D_1^2+D_2^2+D_3^2\right)}\left(a+\sqrt{b}\right)\nonumber\\
\lambda_2 &=&\frac{1}{2 \left(D_1^2+D_2^2+D_3^2\right)}\left(a-\sqrt{b}\right) \ , 
\end{eqnarray}
where
\begin{eqnarray}
a&=& D \Big(D_{11} (D_2^2+ D_3^2) + D_{22} (D_1^2 + D_3^2) + D_{33} (D_1^2 + D_2^2) 
 \cr &-&  2 \left( D_1 D_2 D_{12} + D_2 D_3 D_{23} + D_3 D_1 D_{31}\right)\Big) \ ,  \cr
b&=& a^2 +  4 D^2 (D1^2 + D2^2 + D3^2) f(D_{ab}, D_a) \ , 
\end{eqnarray}
with
\begin{eqnarray}
f(D_{ab},D_a) &= &\Big(
D_1^2 (D_{23}^2 - D_{22}D_{33}) 
+ D_2^2 (D_{13}^2 - D_{11} D_{33} ) 
+  D_3^3 (D_{12}^2 - D_{11} D_{22}) \cr
&+& 2 D_1 D_2 (D_{12} D_{33} - D_{13} D_{23})
+ 2 D_2 D_3 ( D_{11} D_{23} - D_{12} D_{13})  \cr
&+& 2 D_1 D_3 (D_{22} D_{13} - D_{12} D_{23})
\Big) \ . 
\end{eqnarray}
We can simplify this expression. We find $f(D_{ab},D_a) = - D\ {\rm det}(D_{ab})$. Thus, we will have
$a^2 - b > 0$ provided $D \ {\rm det}(D_{ab}) > 0$. However, this is not sufficient to make both $\lambda_1$
and $\lambda_2$ positive. We need, in addition $a>0$. Note that 
\begin{eqnarray*}
 a &=& D\Big( (D_{11} + D_{22} + D_{23}) (D_1^2 + D_2^2 + D_3^2) 
 - D_{11} D_1^2 - D_{22} D_2^2 - D_{33} D_3^2
 \cr &-&  2 \left( D_1 D_2 D_{12} + D_2 D_3 D_{23} + D_3 D_1 D_{31}\right)\Big) \ . 
 \end{eqnarray*}
 The term in parenthesis is nothing but the difference between the product of traces and the trace of 
 products of the matrices $D_{ab}$  and $D_a D_b$:%. In otherwords,
 \begin{eqnarray*}
 a = D\ {\rm Tr}(D_{ab}) {\rm Tr}(D_aD_b) - D\ {\rm Tr}(D_{ab} D_bD_c) \ . 
 \end{eqnarray*}
 So, in the three parameter model in addition to the condition $D\ {\rm det}(D_{ab})>0$ we need
 $D\ {\rm Tr}(D_{ab}) {\rm Tr}(D_aD_b) > D\ {\rm Tr}(D_{ab} D_bD_c)$ in order to get a stable, 
 non-supersymmetric attractor. More genrally, for a $n$-parameter model, we need to find the 
 eigenvalues of the mass term (\ref{leadingmass}). For all of them to be positive we need to 
 impose $(n-1)$ constraints relating the triple intersection numbers $D_{abc}$ and the black hole
 charges. For a generic Calabi-Yau compactification these constraints can always be met by 
 suitable choices of black hole charges. Thus for the $D0-D4$ configuration, we can always have 
 stable non-supersymmetric attractors in a subspace of the charge lattice.
 
\section{Conclusion}

In this paper we analyzed the stability of non-supersymmetric attractors in the presence of stringy 
corrections to the $N=2$ pre-potential. Holomorphicity constrains the pre-potential to take the form 
(\ref{fullprep}) for perturbative corrections to all orders. We considered $D0-D6$ as well as $D0-D4$
configurations in presence of such corrections. We found for the $D0-D6$ system the space of 
attractor points define a $(n-1)$ dimensional hypersurface in the moduli space. The perturbative 
corrections only deform this hypersurface instead of lifting any of the flat directions present at 
the leading order. Interestingly, the $D0-D4$ system behaves differently. In this case, the corrections 
make all the massless modes massive. In the case of two and three parameter models we have 
explicitly obtained the eigenvalues. We saw that the non-supersymmetric attractors are not stable 
at an arbitrary point in the charge lattice. However they are stable in a subspace of the charge lattice.
Interestingly, in both these cases, though the eigenvalues are explicitly dependent on the values of 
the scalar fields at the attractor point, the stability conditions are independent of it.
In the case where we have more than three K\"ahler moduli, it is much harder to diagonalize the 
mass matrix. However we argued that in such cases also there must exist a subspace of the charge
lattice which admit stable non-supersymmetric attractors. We believe the same result will also 
hold for $D0-D4-D6$ configuration.

In this paper we considered intersecting $D$-brane configurations in the 
weak curvature limit and our main focus was on the effect of the $\alpha$-corrections to the $N=2$
pre-potential of the $4D$ supergravity theory. We saw that for the extremal non-supersymmetric black 
holes, these corrections lift all the vector moduli in certain subspace of the charge lattice. Our entire
focus was on the vector multiplet moduli only. The hypermultiplet moduli are still not fixed at the 
black hole horizon. It would be interesting to find a mechanism to fix these hypermultiplet moduli.
This will lead to a better understanding of the microscopic origin of such extremal non-supersymmetric 
black holes.

\section{Acknowledements} 

This work was partially supported by the  IFCPAR/CEFIPRA project number 4104-2. We have been 
benefitted by discussions with Suresh Govindarajan, I. Karthik and Sandip Trivedi  at various times 
during this project. PKT acknowledges kind hospitality by the TIFR (Mumbai), the Ecole Polytechnique and 
the Ecole Normale Superieure (Paris) where parts of this project were carried out. He also
acknowledges the stimulating environment and the hospitality provided by the 
ENS Summer Institute, Paris.

%..........................................................................................................%
%...........................................................................................................%

\appendix

\section{Computational details for the $D0-D6$ system}

In this appendix we will provide some of the computational details for the $D0-D6$ system. The superpotential 
for the system is given by:
\begin{equation}
W=q-p^{0}\hat{\alpha}_{0a}x^{a}+p^{0}D_{abc}x^{a}x^{b}x^{c} \ . 
\end{equation}
In this section we denote $q = q_0 - 2 i p^0 \beta$. We are interested in analyzing the attractor equation
\begin{equation}\label{veffprim2}
 \left(g^{b\bar c} \nabla_a\nabla_b W \overline{\nabla_cW} + 2 \nabla_a W \overline{W}
+ \partial_a g^{b\bar c} \nabla_b W\overline{\nabla_c W}\right) = 0~.
\end{equation}
In the following we will evaluate each of the terms separately. 
The covariant derivative of W takes the form:
\begin{equation}
\nabla_{a}W=3p^{0}D_{abc}x^{b}x^{c}-p^{0}\hat{\alpha}_{0a}-\frac{3M_{a}W}{M-4 i\beta} \ , 
\end{equation}
where as the double derivative term becomes
\begin{eqnarray}
\nabla_{a}\nabla_{b}W&=&6p^{0}D_{abp}x^{p}+\frac{6W}{M- 4 i \beta}\left(\frac{3M_{a}M_{b}}{M- 4 i \beta}-M_{ab}\right)\nonumber\\
&+&\frac{3p^{0}}{M- 4 i \beta}\left(M_{a}\hat{\alpha_{0}}_{b}+M_{b}\hat{\alpha_{0}}_{a}\right)
-\frac{9p^{0}x^{p}x^{q}}{M- 4 i \beta}\left(M_{a}D_{bpq}+M_{b}D_{apq}\right) \ .
\end{eqnarray}
We use the ansatz $x^{a}=\hat{x}^{a}(t_{1}+it_{2})$ and simplify to get  
\begin{equation}
\nabla_{a}W=\hat{D_{a}}\left(3t^{2}p^{0}+\frac{3iW}{2Dt_{2}(1+s)}\right)-p^{0}\hat{\alpha}_{0a} \ , 
\label{susy}
\end{equation}
where $\hat{D_{a}}=D_{{abc}}\hat{x}^{b}\hat{x}^{c}$, $\hat{D}=\hat{D_{a}} \hat{x}^{a}$ and $s= %\frac{L}{8\hat{D}{t_{2}}^{3}}=
{\beta}/{(2\hat{D}{t_{2}}^{3})}$. 
The double derivative term in this ansatz becomes:
\begin{eqnarray}
\nabla_{a}\nabla_{b}W&=&3\hat{D_{ab}}\left(2p^{0}t+\frac{W}{2\hat{D}{t_{2}}^{2}(1+s)}\right)-\frac{3ip^{0}}{2t_{2}\hat{D}(1+s)}\left(\hat{D_{a}}\hat{\alpha_{0}}_{b}+\hat{D_{b}}\hat{\alpha_{0}}_{a}\right)\nonumber\\&+&\frac{9\hat{D_{a}}\hat{D_{b}}}{\hat{D}}\left(\frac{-W}{2\hat{D}{t_{2}}^{2}(1+s)^{2}}+\frac{it^{2}p^{0}}{t_{2}(1+s)}\right).
\end{eqnarray}
We use the following notations
\begin{eqnarray}
\nabla_aW&=&a_1\hat{D_a}+a_2\hat{\alpha_{0}}_a\nonumber\\
\nabla_{a}\nabla_{b}W&=&a_{3}\hat{D_{ab}}+a_{4}\frac{\hat{D_{a}}\hat{D_{b}}}{\hat{D}}+a_5\left(\hat{D_a}\hat{\alpha_{0}}_b+\hat{D_b}\hat{\alpha_{0}}_a\right)\nonumber\\
g^{b\bar{c}}&=&a_6\hat{x}^b\hat{x}^c+a_7\hat{D}\hat{D}^{bc} \ , 
\label{ddw}
\end{eqnarray}
where
\begin{eqnarray}
a_1&=&3\left(p^{0}t^2+\frac{iW}{2\hat{D}t_2(1+s)}\right)\nonumber\\
a_2&=&-p^{0}\nonumber\\
 a_{3}&=&3\left(2p^{0}t+\frac{W}{2\hat{D}{t_{2}}^{2}(1+s)}\right)\nonumber\\
 a_{4}&=&\frac{9}{t_{2}(1+s)}\left(ip^{0}t^{2}-\frac{W}{2t_{2}D(1+s)}\right)\nonumber
 \end{eqnarray} \begin{eqnarray}
a_5&=&\frac{-3ip^{0}}{2\hat{D}t_{2}(1+s)}\nonumber\\
a_6&=&2t_2^2\frac{(1+s)}{1-2s}\nonumber\\
a_7&=&\frac{-2t_2^2(1+s)}{3}\nonumber\\
a_8&=&\bar{a_1}a_6\hat{D}+\bar{a_1}a_7\hat{D}+\bar{a_2}a_6\hat{\alpha_{0}}\nonumber\\
a_9&=&\bar{a_2}a_7\nonumber
\end{eqnarray}
  Now we list the three terms in the equation of motion
\begin{eqnarray}
g^{b\bar{c}}\nabla_{a}\nabla_{b}W\overline{\nabla_{c}W}&=&\hat{D_a}\left(a_8a_3+a_8a_4+a_8a_5\hat{\alpha_{0}}+a_9a_4\hat{\alpha_{0}}+a_9a_5\hat{D}T\right)\nonumber\\&+&\hat{\alpha_{0}}_a\left(\hat{D}(a_8a_5+a_9a_3+a_9a_5\hat{\alpha_{0}})\right)\nonumber\\
\partial_{a}g^{b\bar{c}}\nabla_{b}W\overline{\nabla_{c}W}&=&it_2\hat{D}_a\left(|a_1|^2\hat{D}\left(1+\frac{(2s-7)(1+s)}{3(1-2s)}+\frac{9s}{(1-2s)^2}\right)\right.\nonumber\\&+&\left.(\bar{a_1}a_2+\bar{a_2}a_1)\hat{\alpha_{0}}\left(1-\frac{1+s}{1-2s}+\frac{9s}{(1-2s)^2}\right)\right.\nonumber\\&+&\left.|a_2|^2\left(T+\frac{9s\hat{\alpha_{0}}^2}{\hat{D}(1-2s)^2}\right)\right)-it_2|a_2|^2\hat{D}\frac{(1+s)}{3}T_a
\nonumber\\&+&2it_2\frac{(1+s)}{(1-2s)}\hat{\alpha_{0}}_a\left(\frac{\hat{D}(s-2)}{3}(\bar{a_2}a_1+\bar{a_1}a_2)-|a_2|^2\hat{\alpha_{0}}\right)\nonumber\\
2\nabla_{a}W\bar{W}&=&2\overline{W}\left(a_1\hat{D}_a+a_2\hat{\alpha_{0}}_a\right)
\end{eqnarray}
We introduced the notations $T=p^{0}\hat{D}^{bc}\hat{\alpha_{0}}_b\hat{\alpha_{0}}_c$ and 
$T_a=p^{0}D_{apq}\hat{D}^{pb}\hat{D}^{qc}\hat{\alpha_{0}}_b\hat{\alpha_{0}}_c$. 
Adding all the terms in the above equation we get the equation of motion corresponding to the non-supersymmetric solution. Taking the real part of the equation of motion we get
\begin{equation}
a_{10}\hat{D}_a-a_{11}{\alpha_{0}}_a=0 \ , 
\end{equation}
where
\begin{eqnarray}
a_{10}&=&3 \left(4p^{0} \hat{D}{t_1}{t_2}^4 \left(p^{0}\left({t_1}^2+{t_2}^2\right) \left(\hat{D} \left({t_1}^2+{t_2}^2\right)-\hat{\alpha_{0}}
   \right)+{q_0}{t_1}\right)\right.\nonumber\\&+&\left.{t_2} p^{0}\beta  \left(-p^{0}\hat{D} \left({t_1}^5-2{t_1}^3{t_2}^2+5{t_1}
   {t_2}^4\right)\right.\right.\nonumber\\&-&\left.\left.{q_0} \left({t_1}^2-3{t_2}^2\right)+{t_1}p^{0}\hat{\alpha_{0}} \left({t_1}^2+{t_2}^2\right)\right)+2{t_1}
  (p^{0})^{2} \beta ^2 \left({t_1}^2+5{t_2}^2\right)\right)\nonumber\\
a_{11}&=&-{t_2}p^{0}\beta \left(p^{0}\hat{D}{t_1}^3-11p^{0}\hat{D}{t_1}
   {t_2}^2+{q_0}-{t_1}p^{0} \hat{\alpha_{0}} \right)\nonumber\\&-&4p^{0}\hat{ D}
   {t_2}^4 \left({t_1} \left(p^{0}\hat{D}
   \left({t_2}^2-{t_1}^2\right)+p^{0}\hat{\alpha_{0}}
   \right)-{q_0}\right)+2{t_1}(p^{0})^{2} \beta ^2\nonumber \ . 
\end{eqnarray}

At the critical point
\begin{equation}
\hat{\alpha_{0}}_a=G\hat{D}_a=\frac{a_{10}}{a_{11}}\hat{D}_a
\end{equation}
Contracting with $\hat x^a$ on both sides of the above equation we get
\begin{equation} \label{modulispace}
G=\frac{\hat{\alpha_{0}}}{\hat{D}}
\end{equation}
For convenience we introduce the following rescaling: 
\begin{eqnarray}
t_{1}&=&\frac{xq_{0}}{p^{0}\hat{\alpha}_{0}}\\
t_{2}&=&y\sqrt{\hat{\alpha}_{0}/\hat{D}}\\
\hat{D}&=&\tilde{D} \frac{\hat{\alpha}_{0}^{3}(p^{0})^{2}}{q_{0}^{2}}\\
\beta&=&\frac{\gamma q_{0}}{p^{0}\tilde{D}^{{1/2}}}
\end{eqnarray}
Taking the original equation of motion, contract it with $\hat{x}^a$. Separate the real and imaginary part and using the above change of variables we get the equations of motion for the D0-D6 system:
   \begin{eqnarray}
0&=&-y \gamma  \left(x \left(3\tilde{ D}^2 x^4+\tilde{D} x \left(3-2 x \left(3 y^2+2\right)\right)+15 y^4+8 y^2+1\right)-9 y^2-1\right)\nonumber\\&+&4 y^4 \left(x \left(3 \tilde{D}^2 x^4+\tilde{D }x \left(x
   \left(6 y^2-4\right)+3\right)+3 y^4-2 y^2+1\right)-1\right)\nonumber\\&+&2 x \gamma ^2 \left(3 \tilde{D} x^2+15 y^2-1\right)
      \end{eqnarray} and
      
              \begin{eqnarray}
      0&=&2 y^3 \gamma ^3 \left(9 \tilde{D}^2 x^4+6 \tilde{D} x \left(x \left(30
   y^2+2\right)-9\right)+99 y^4-54 y^2+7\right)\nonumber\\&+&\gamma ^4 \left(18 y^2
   \left(1-5 \tilde{D} x^2\right)+\left(1-3 \tilde{D} x^2\right)^2-63 y^4\right)\nonumber\\&-&3 y^4
   \gamma ^2 \left(y^2 \left(3 \tilde{D} x \left(33 \tilde{D} x^3-20
   x-6\right)+13\right)+15 \tilde{D} \left(\tilde{D} x^3-x+1\right)^2\right.\nonumber\\&+&\left.9 y^4 \left(41 \tilde{D}
   x^2-6\right)+45 y^6\right)+8 y^{10} \left(-9 \tilde{D} \left(\tilde{D}
   x^3-x+1\right)^2+9 \tilde{D} x^2 y^4\right.\nonumber\\&-&\left.y^2 \left(1-3 \tilde{D} x^2\right)^2+9
   y^6\right)+4 y^7 \gamma  \left(4 y^2 \left(2-3 \tilde{D} x \left(3 \tilde{D} x^3-5
   x+9\right)\right)\right.\nonumber\\&+&\left.9 \tilde{D} \left(\tilde{D} x^3-x+1\right)^2+9 y^4 \left(\tilde{D}
   x^2-2\right)-18 y^6\right)   \end{eqnarray}
   Substituting the solution of the above two equations in Eq.(\ref{modulispace}) we get the $(n-1)$ dimensional
   hypersurface  defining the space of attractor points.
  %%%%%%%%%%%%%%%%%%%%%%%%%%%%%%%%%%%%
\section{Computational details for the $D0-D4$ system}
In this appendix we will outline some of the computational details needed for the $D0-D4$ system. The 
superpotential: 
\begin{eqnarray}
W=q_0 - \alpha_{0a} p^a -3D_{ab}x^{a}x^{b} \ .
\end{eqnarray}
and its covariant derivatives with ansatz $x^a = p^a (t_1 + i t_2)$ are given by 
\begin{eqnarray*} 
\nabla_{a}W&=&3D_{a}\left(\frac{iW}{2t_{2}D(1+s)}-2t\right)\cr
\nabla_{a}\nabla_{b}W&=&3D_{ab}\left(-2+\frac{W}{2{t_{2}}^{2}D(1+s)}\right)-\frac{9D_{a}D_{b}}{Dt_{2}(1+s)}\left(2it+\frac{W}{2t_{2}D(1+s)}\right) 
\end{eqnarray*}
The mass matrix is given by the double derivative terms of the effective potential. It is straightforward to see 
that 
\begin{eqnarray}
\label{atwoderiv}
e^{-K_0} \partial_a\partial_d V &=& \left\{ 
g^{b\bar c} \nabla_a\nabla_b\nabla_d W 
+ \partial_a g^{b\bar c} \nabla_b\nabla_d W + \partial_d g^{b\bar c} \nabla_b\nabla_a W
\right\} \overline{\nabla_cW} \cr &+&
 3 \nabla_a\nabla_d W \overline{W}
+ \partial_a\partial_d g^{b\bar c} 
\nabla_b W \overline{\nabla_cW}
- g^{b\bar c}\partial_a g_{d\bar c} \nabla_b W \overline{W} \cr
e^{-K_0} \partial_a\partial_{\bar d} V &=&
g^{b\bar c} \nabla_a\nabla_b W\overline{\nabla_c \nabla_d W}
+ \left\{ 2 |W|^2 + g^{b\bar c}  \nabla_b W\overline{\nabla_c W}\right\} g_{a\bar d}
\cr &+&
 \partial_a g^{b\bar c} \nabla_b W \overline{\nabla_c \nabla_d W}
+ \partial_{\bar d} g^{b\bar c} \nabla_a\nabla_b W \overline{\nabla_c W}
+ 3 \nabla_a W\overline{\nabla_d W}
\cr &+&
  \partial_a\partial_{\bar d} g^{b\bar c} \nabla_b W \overline{\nabla_c W}
\end{eqnarray}
We need to evaluate each of the terms in the above equation at the attractor point. After some tedious 
algebra we find 
\begin{eqnarray*}
g^{b\bar c} \nabla_a\nabla_b W \overline{\nabla_c\nabla_dW} &=& {D}_a{D}_d\left(\bar{g_2}g_3+\bar{g_1}g_3+\bar{g_2}g_4\right)+\bar{g_1}g_4{D}{D}_{ad}\cr
3 \nabla_aW\overline{\nabla_dW} &=& 3|h|^{2}{D}_a{D}_d\cr
2 g_{a\bar d} |W|^2 &=&-\frac{3W\overline{W}}{2D(1+s){t_{2}}^{2}}\left(2D_{ad}-\frac{3D_{a}D_{d}}{D(1+s)}\right)\cr
g_{a\bar d} g^{b\bar c} \nabla_bW \overline{\nabla_cW} &=&-\frac{|h|^{2}D(1+s)}{1-2s}\left(2D_{ad}-\frac{3D_{a}D_{d}}{D(1+s)}\right)\cr
\partial_a g^{b\bar c} \nabla_bW \overline{\nabla_c\nabla_dW} &=& DD_{ad}\bar{g_{1}}g_{5}+D_{a}D_{d}\left(\bar{g_{2}}g_{5}+\bar{g_{1}}g_{6}+\bar{g_{2}}g_{6}\right)\cr
\partial_{\bar d} g^{b\bar c} \overline{\nabla_cW} \nabla_a\nabla_dW &=&DD_{ad}{g_{1}}\bar{g_{5}}+D_{a}D_{d}\left({g_{2}}\bar{g_{5}}+{g_{1}}\bar{g_{6}}+{g_{2}}\bar{g_{6}}\right)\cr
\partial_a\partial_{\bar d} g^{b\bar c} \nabla_bW \overline{\nabla_cW} &=& DD_{ad}(1+s)|h|^{2}\left(\frac{2}{1-2s}-\frac{1}{3}-\frac{3}{(1-2s)^{2}}\right) \cr
&+&D_{a}D_{d}|h|^{2}\left(1+\frac{7+s}{1-2s}-\frac{3(5+2s)}{(1-2s)^{2}}+\frac{9(1+s)}{(1-2s)^{3}}\right)\cr
\end{eqnarray*}
and
\begin{eqnarray*}
g^{b\bar c} \nabla_a\nabla_b\nabla_d W \overline{\nabla_cW} &=&\left(D_{ad}g_9+D_aD_dg_{10}\right)\frac{4\bar{h}D{t_2}^2(1+s)^2}{3(1-2s)}\cr
\partial_ag^{b\bar{c}}\nabla_b\nabla_dW\overline{\nabla_cW}&=&DD_{ad}\left(g_7g_1\right)+D_aD_d\left(g_7g_2+g_8g_1+g_8g_2\right)\cr
3 \nabla_a\nabla_dW\overline{W} &=&3\overline{W}\left(g_1D_{ad}+g_2\frac{D_aD_d}{D}\right)\cr
- g^{b\bar c} \partial_a g_{d\bar c} \nabla_bW\overline{W} &=&-\frac{i\overline{W}h}{t_2(1-2s)}\left(D_{ad}(s-2)+3\frac{D_aD_d}{D}\frac{(1-2s)}{(1+s)}\right)\cr
\partial_a\partial_dg^{b\bar c}\nabla_bW\overline{\nabla_cW} &=&D(1+s)|h|^2D_{ad}\left(\frac{1}{3}-\frac{2}{1-2s}+\frac{3}{(1-2s)^2}\right)\cr
&+&|h|^2D_aD_d\left(-1-\frac{(7+s)}{(1-2s)}+\frac{3(5+2s)}{(1-2s)^2}-\frac{9(1+s)}{(1-2s)^3}\right)
\end{eqnarray*}
We used the following notations in the above series of equations
\begin{eqnarray}
g_1&=&3\left(-2+\frac{W}{2D{t_2}^2(1+s)}\right)\nonumber\\
g_2&=&\frac{-9}{t_{2}(1+s)}\left(2it+\frac{W}{2D{t_2}(1+s)}\right)\nonumber\\
g_3&=&\frac{2{t_2}^2(1+s)}{3}\left(\frac{3(g_1+g_2)}{1-2s}-g_2\right)\nonumber
\end{eqnarray} \begin{eqnarray}
g_4&=&-\frac{2{t_2}^2(1+s)g_1}{3}\nonumber\\
g_5&=&2iht_2\frac{(s-2)(1+s)}{3(1-2s)}\nonumber\\
g_6&=&iht_2\left(1-\frac{(1+s)}{(1-2s)}+\frac{9s}{(1-2s)^2}\right)\nonumber\\
g_7&=&2it_2\bar{h}\frac{(s-2)(1+s)}{3(1-2s)}\nonumber\\
g_8&=&it_2\bar{h}\left(1+\frac{9s}{(1-2s)^2}-\frac{1+s}{1-2s}\right)\nonumber\\
g_9&=&\frac{-3}{t_{2}(1+s)}\left(3i+\frac{3t}{{t_2}}+\frac{iW}{4{t_2}^2D}-\frac{6iW}{4{t_2}^2D(1+s)}\right)\nonumber\\
g_{10}&=&\frac{3}{t_{2}(1+s)}\left(-3+\frac{i(g_{1}+g_{2})}{2}-\frac{9t}{t_{2}(1+s)}-\frac{3t}{t_{2}}\right.\nonumber\\&+&\left.\frac{J(s-2)}{2t_{2}(1+s)}+\frac{3iW(s-5)}{4D{t_{2}}^{2}(1+s)^{2}}+\frac{6iW}{4D{t_{2}}^{2}(1+s)}\right) \nonumber
\end{eqnarray}
The double derivative terms of the effective potential can now be written as 
\begin{eqnarray} \label{frstddv}
e^{-K_{0}}\partial_a\partial_{\bar{d}}V&=&\frac{3q}{2 D y \left(\gamma -2 y^3\right)^2 \left(y^3+\gamma \right)^3}\left(DD_{ad}X_1+D_aD_dX_2\right)\\
e^{-K_{0}}\partial_a\partial_{d}V&=&\frac{-3 q \left(y^3+y-2 \gamma \right)}{2 D y \left(2 y^3-\gamma \right)^2 \left(y^3+\gamma \right)^3}\left(DD_{ad}Y_1+D_aD_dY_2\right)\label{sndddv}
\end{eqnarray}
where we have defined 
\begin{eqnarray}
X_1&=&80 y^{18}-32 y^{16}+144 y^{15} \gamma +16 y^{14}-160 y^{13}
   \gamma +150 y^{12} \gamma ^2\nonumber\\&+&16 y^{11} \gamma -68 y^{10}
   \gamma ^2-23 y^9 \gamma ^3+6 y^8 \gamma ^2+58 y^7 \gamma
   ^3-81 y^6 \gamma ^4\nonumber\\&+&y^5 \gamma ^3+2 y^4 \gamma ^4+24 y^3
   \gamma ^5-5 y^2 \gamma ^4+4 y \gamma ^5-4 \gamma ^6\\
X_2&=&-144 y^{18}-180 y^{15} \gamma -48 y^{14}+360 y^{13} \gamma -522
   y^{12} \gamma ^2-36 y^{11} \gamma \nonumber\\&+&108 y^{10} \gamma ^2+99
   y^9 \gamma ^3-90 y^8 \gamma ^2+270 y^7 \gamma ^3-180 y^6
   \gamma ^4-21 y^5 \gamma ^3\nonumber\\&+&36 y^4 \gamma ^4-36 y^3 \gamma ^5\\
Y_1&=&16 y^{15}+16 y^{13}-64 y^{12} \gamma +16 y^{10} \gamma -74 y^9
   \gamma ^2+6 y^7 \gamma ^2+25 y^6 \gamma ^3\nonumber\\&+&y^4 \gamma ^3+17
   y^3 \gamma ^4-5 y \gamma ^4-2 \gamma ^5\\
Y_2&=&48 y^{15}-48 y^{13}+300 y^{12} \gamma -36 y^{10} \gamma +126 y^9
   \gamma ^2-90 y^7 \gamma ^2\nonumber\\&+&111 y^6 \gamma ^3-21 y^4 \gamma
   ^3-6 y^3 \gamma ^4
\end{eqnarray}
We will use the rescale variables
\begin{eqnarray}
y&=&\omega z\nonumber\\
\gamma&=&\omega^3
\end{eqnarray}
and the equation of motion in terms of them
\begin{equation}
8 z^9 \omega ^2-8 z^7+36 z^6 \omega ^2-4 z^4+3 z^3 \omega ^2-5 z+2 \omega ^2=0
\end{equation}
Using the equation of motion we can rewrite Eqs.(\ref{frstddv}) and (\ref{sndddv}) as 
\begin{eqnarray}
e^{-K_{0}}\partial_a\partial_dV&=&A_{1}D_{ad}+A_{2}\frac{D_{a}D_{d}}{D}\nonumber\\
e^{-K_{0}}\partial_a\partial_{\bar{d}}V&=&\frac{3D_aD_d}{D}E_1-D_{ad}E_2 \ , 
\end{eqnarray}
or, in other words, the mass matrix has the form
\begin{eqnarray*}
M=\left(\frac{3D_aD_d}{D}E_1(z_0)-D_{ad}E_2(z_0)\right)\otimes I
+\left(A_1(z_0)D_{ad}+A_2(z_0)\frac{D_aD_d}{D}\right)\otimes\sigma^3  , \ \ \ 
\end{eqnarray*}where the functions $E_1(z), E_2(z), A_1(z)$ and $A_2(z)$ are given by 
\begin{eqnarray}
E_1&=&-e^{K_{0}}\frac{24 q z^3 \left(4 z^3+1\right) \left(16 z^9+24 z^3+13\right)}{\left(8 z^6+4 z^3+5\right) \left(8 z^9+36 z^6+3 z^3+2\right)}\nonumber\\
E_2&=&-e^{K_{0}}\frac{24 q \left(2 z^3-1\right)^3}{8 z^9+36 z^6+3 z^3+2}\nonumber\\
A_1&=&-e^{K_{0}}\frac{48 q \left(1-2 z^3\right)^2 \left(z^3+1\right)}{8 z^9+36 z^6+3 z^3+2}\nonumber\\
A_2&=&-e^{K_{0}}\frac{864 q z^3 \left(4 z^9-3 z^3+1\right)}{\left(8 z^6+4 z^3+5\right) \left(8 z^9+36 z^6+3 z^3+2\right)}
\end{eqnarray}
The mass matrix can be written in the block diagonalized form
\begin{eqnarray}\label{blockdiag}
e^{-K_0} M&=& - \frac{24q}{D\left(8z_0^9+36z_0^6+3z_0^3+2\right)}\left(
\begin{array}{cc}
 M_t(z_0)& 0 \\
 0 &M_b(z_0)
\end{array}
\right) \ ,
\end{eqnarray}
Where
\begin{eqnarray}
M_t(z)&=&\frac{3D_aD_d\left(16\left(4z^9-2z^6+6z^3+ 7\right)z^6+z^3\right)}{8z^6+4z^3+5}-DD_{ad} \left(4 z^3+1\right)\left(1-2z^3\right)^2 \cr
M_b(z)&= &3\left(DD_{ad}\left(1-2z^3\right)^2+D_aD_d\left(8z^6+4z^3+5\right)z^3\right) \ .
\label{matrixmb}
\end{eqnarray}

%%%%%%%
%%%%%%%%
%%%%%%


\begin{thebibliography}{99}
%\cite{Ferrara:1995ih}
\bibitem{Ferrara:1995ih}
  S.~Ferrara, R.~Kallosh, A.~Strominger,
  %``N=2 extremal black holes,''
  Phys.\ Rev.\  {\bf D52 } (1995)  5412-5416.
  [hep-th/9508072].

%\cite{Ferrara:1997tw}
\bibitem{Ferrara:1997tw}
  S.~Ferrara, G.~W.~Gibbons, R.~Kallosh,
  %``Black holes and critical points in moduli space,''
  Nucl.\ Phys.\  {\bf B500}, 75-93 (1997).
  [hep-th/9702103].

%\cite{Goldstein:2005hq}
\bibitem{Goldstein:2005hq}
  K.~Goldstein, N.~Iizuka, R.~P.~Jena, S.~P.~Trivedi,
  %``Non-supersymmetric attractors,''
  Phys.\ Rev.\  {\bf D72}, 124021 (2005).
  [hep-th/0507096].

%\cite{Strominger:1996kf}
\bibitem{Strominger:1996kf}
  A.~Strominger,
  %``Macroscopic entropy of N=2 extremal black holes,''
  Phys.\ Lett.\  {\bf B383 } (1996)  39-43.
  [hep-th/9602111].

%\cite{Ferrara:1996dd}
\bibitem{Ferrara:1996dd}
  S.~Ferrara, R.~Kallosh,
  %``Supersymmetry and attractors,''
  Phys.\ Rev.\  {\bf D54}, 1514-1524 (1996).
  [hep-th/9602136].

%\cite{Ferrara:1996um}
\bibitem{Ferrara:1996um}
  S.~Ferrara, R.~Kallosh,
  %``Universality of supersymmetric attractors,''
  Phys.\ Rev.\  {\bf D54}, 1525-1534 (1996).
  [hep-th/9603090].

%\cite{Behrndt:1996jn}
\bibitem{Behrndt:1996jn}
  K.~Behrndt, G.~Lopes Cardoso, B.~de Wit, R.~Kallosh, D.~Lust, T.~Mohaupt,
  %``Classical and quantum N=2 supersymmetric black holes,''
  Nucl.\ Phys.\  {\bf B488}, 236-260 (1997).
  [hep-th/9610105].


%\cite{Behrndt:1998eq}
\bibitem{Behrndt:1998eq}
  K.~Behrndt, G.~Lopes Cardoso, B.~de Wit, D.~Lust, T.~Mohaupt, W.~A.~Sabra,
  %``Higher order black hole solutions in N=2 supergravity and Calabi-Yau string backgrounds,''
  Phys.\ Lett.\  {\bf B429}, 289-296 (1998).
  [hep-th/9801081].

%\cite{LopesCardoso:1998wt}
\bibitem{LopesCardoso:1998wt}
  G.~Lopes Cardoso, B.~de Wit, T.~Mohaupt,
  %``Corrections to macroscopic supersymmetric black hole entropy,''
  Phys.\ Lett.\  {\bf B451}, 309-316 (1999).
  [hep-th/9812082].


%\cite{Ooguri:2004zv}
\bibitem{Ooguri:2004zv}
  H.~Ooguri, A.~Strominger, C.~Vafa,
  %``Black hole attractors and the topological string,''
  Phys.\ Rev.\  {\bf D70}, 106007 (2004).
  [hep-th/0405146].


%\cite{Strominger:1996sh}
\bibitem{Strominger:1996sh}
  A.~Strominger, C.~Vafa,
  %``Microscopic origin of the Bekenstein-Hawking entropy,''
  Phys.\ Lett.\  {\bf B379}, 99-104 (1996).
  [hep-th/9601029].

%\cite{David:2002wn}
\bibitem{David:2002wn}
  J.~R.~David, G.~Mandal, S.~R.~Wadia,
  %``Microscopic formulation of black holes in string theory,''
  Phys.\ Rept.\  {\bf 369}, 549-686 (2002).
  [hep-th/0203048].

%\cite{Mohaupt:2008gt}
\bibitem{Mohaupt:2008gt}
  T.~Mohaupt,
  %``From Special Geometry to Black Hole Partition Functions,''
  
  [arXiv:0812.4239 [hep-th]].

%\cite{Bellucci:2007ds}
\bibitem{Bellucci:2007ds}
  S.~Bellucci, S.~Ferrara, R.~Kallosh, A.~Marrani,
  %``Extremal Black Hole and Flux Vacua Attractors,''
  Lect.\ Notes Phys.\  {\bf 755}, 115-191 (2008).
  [arXiv:0711.4547 [hep-th]].

%\cite{Bellucci:2008jq}
\bibitem{Bellucci:2008jq}
  S.~Bellucci, S.~Ferrara, A.~Marrani,
  %``Attractors in Black,''
  Fortsch.\ Phys.\  {\bf 56}, 761-785 (2008).
  [arXiv:0805.1310 [hep-th]].

%\cite{Ceresole:2007wx}
\bibitem{Ceresole:2007wx}
  A.~Ceresole, G.~Dall'Agata,
  %``Flow Equations for Non-BPS Extremal Black Holes,''
  JHEP {\bf 0703}, 110 (2007).
  [hep-th/0702088].


%\cite{Tripathy:2005qp}
\bibitem{Tripathy:2005qp}
  P.~K.~Tripathy, S.~P.~Trivedi,
  %``Non-supersymmetric attractors in string theory,''
  JHEP {\bf 0603}, 022 (2006).
  [hep-th/0511117].

%\cite{Nampuri:2007gv}
\bibitem{Nampuri:2007gv}
  S.~Nampuri, P.~K.~Tripathy, S.~P.~Trivedi,
  %``On The Stability of Non-Supersymmetric Attractors in String Theory,''
  JHEP {\bf 0708}, 054 (2007).
  [arXiv:0705.4554 [hep-th]].
  
  
%\cite{Cecotti:1988qn}
\bibitem{Cecotti:1988qn}
  S.~Cecotti, S.~Ferrara, L.~Girardello,
  %``Geometry of Type II Superstrings and the Moduli of Superconformal Field Theories,''
  Int.\ J.\ Mod.\ Phys.\  {\bf A4}, 2475 (1989).
  
  %\cite{Hosono:1993qy}
\bibitem{Hosono:1993qy}
  S.~Hosono, A.~Klemm, S.~Theisen, S.~-T.~Yau,
  %``Mirror symmetry, mirror map and applications to Calabi-Yau hypersurfaces,''
  Commun.\ Math.\ Phys.\  {\bf 167}, 301-350 (1995).
  [hep-th/9308122].
  
%\cite{Hosono:1994ax}
\bibitem{Hosono:1994ax}
  S.~Hosono, A.~Klemm, S.~Theisen, S.~-T.~Yau,
  %``Mirror symmetry, mirror map and applications to complete intersection Calabi-Yau spaces,''
  Nucl.\ Phys.\  {\bf B433}, 501-554 (1995).
  [hep-th/9406055].

  %\cite{Hosono:1995bm}
\bibitem{Hosono:1995bm}
  S.~Hosono, B.~H.~Lian, S.~-T.~Yau,
  %``GKZ generalized hypergeometric systems in mirror symmetry of Calabi-Yau hypersurfaces,''
  Commun.\ Math.\ Phys.\  {\bf 182}, 535-578 (1996).
  [alg-geom/9511001].

  

%\cite{Becker:2002nn}
\bibitem{Becker:2002nn}
  K.~Becker, M.~Becker, M.~Haack, J.~Louis,
  %``Supersymmetry breaking and alpha-prime corrections to flux induced potentials,''
  JHEP {\bf 0206}, 060 (2002).
  [hep-th/0204254].




%\cite{Dominic:2011iz}
\bibitem{Dominic:2011iz}
  P.~Dominic, P.~K.~Tripathy,
  %``On the Stability of Non-Supersymmetric Quantum Attractors in String Theory,''
  JHEP {\bf 1106}, 112 (2011).
  [arXiv:1105.0481 [hep-th]].

%\cite{Dominic:2010yv}
\bibitem{Dominic:2010yv}
  P.~Dominic, P.~K.~Tripathy,
  %``Instanton Corrected Non-Supersymmetric Attractors,''
  JHEP {\bf 1101}, 116 (2011).
  [arXiv:1010.3373 [hep-th]].
  
  %\cite{Bellucci:2010zd}
\bibitem{Bellucci:2010zd}
  S.~Bellucci, A.~Marrani, R.~Roychowdhury,
  %``Topics in Cubic Special Geometry,''
  
  [arXiv:1011.0705 [hep-th]].


%\cite{Bellucci:2007eh}
\bibitem{Bellucci:2007eh}
  S.~Bellucci, S.~Ferrara, A.~Marrani, A.~Shcherbakov,
  %``Splitting of Attractors in 1-modulus Quantum Corrected Special Geometry,''
  JHEP {\bf 0802}, 088 (2008).
  [arXiv:0710.3559 [hep-th]].

%\cite{Bellucci:2008tx}
\bibitem{Bellucci:2008tx}
  S.~Bellucci, S.~Ferrara, A.~Marrani, A.~Shcherbakov,
  %``Quantum Lift of Non-BPS Flat Directions,''
  Phys.\ Lett.\  {\bf B672}, 77-81 (2009).
  [arXiv:0811.3494 [hep-th]].






\end{thebibliography}
\end{document}